\documentclass[a4paper,fleqn,usenatbib,onecolumn]{mnras}
\usepackage[T1]{fontenc}
\usepackage{ae,aecompl}
\usepackage{graphicx}	
\usepackage{amsmath}	
\usepackage{amssymb}	
\usepackage{natbib}
\usepackage[english]{babel}

\title[The radial dependence of dark matter distribution in M33]{The radial dependence of dark matter distribution in M33}

\author[E. L\'{o}pez Fune et al.]{
E. L\'{o}pez Fune,$^{1}$ $^{2}$\thanks{E-mail: elopez@sissa.it}
P. Salucci,$^{1}$ $^{3}$\thanks{E-mail: salucci@sissa.it}
E. Corbelli,$^{4}$\thanks{E-mail: edvige@arcetri.astro.it}\\
$^{1}$Scuola Internazionale Superiore di Studi Avanzati (SISSA). Via Bonomea, $\#$ 265, 34136 Trieste, Italy.\\
$^{2}$Abdus Salam International Centre for Theoretical Physics. Strada Costiera $\#11$, 34151 Trieste, Italy.\\
$^{3}$INFN, Sezione di Trieste. Via Valerio 2, 34127, Trieste, Italy.\\
$^{4}$INAF-Osservatorio Astrofisico di Arcetri. Largo E. Fermi, 5, 50125 Firenze, Italy.\\
}

\date{Accepted XXX. Received YYY; in original form ZZZ}

\pubyear{2017}

\begin{document}
\label{firstpage}
\pagerange{\pageref{firstpage}--\pageref{lastpage}}
\maketitle

\begin{abstract}
The stellar and gaseous mass distributions, as well as the extended rotation curve in the nearby galaxy M33 are used to derive the radial distribution of dark matter density in the halo and to test cosmological models of galaxy formation and evolution. Two methods are examined to constrain dark mass density profiles. The first method deals directly with fitting the rotation curve data in the range of galactocentric distances $0.24\,\text{kpc}\leq r\leq22.72\,\text{kpc}.$ As found in a previous paper by \citet{Corbelli:2014lga}, and using the results of collisionless $\Lambda-$Cold Dark Matter numerical simulations, we confirm that the Navarro-Frenkel-White (hereafter NFW) dark matter profile provides a better fit to the rotation curve data than the cored Burkert profile (hereafter BRK) profile. The second method relies on the local equation of centrifugal equilibrium and on the rotation curve slope. In the aforementioned range of distances we fit the observed velocity profile, using a function which has a rational dependence on the radius, and derive the slope of the rotation curve. Following \citet{Salucci:2010qr} we then infer the effective matter densities. In the radial range $9.53\,\text{kpc}\leq r\leq22.72\,\text{kpc}$ the uncertainties induced by the luminous matter (stars and gas) becomes negligible, because the dark matter density dominates, and we can determine locally the radial distribution of dark matter. With this second method we tested the NFW and the BRK dark matter profiles and confirm that both profiles are compatible with the data even though in this case the cored BRK density profile provides a more reasonable value for the baryonic-to-dark matter ratio.
\end{abstract}



\section{Introduction}

In the innermost parts of galaxy disks the luminous matter dominates the dynamics: light traces the mass inferred from disk rotation \citep{Athanassoula:1987hp, Persic:1990uj, Palunas:2000qj} out to a radius ranging between 1 and 3 disk exponential scale length (depending upon the galaxy luminosity \citep{Salucci:1999qu}). Instead in the outer skirts, far from galaxy centres, rotational velocities are found to be constant or even rising with radius despite the sharp radial decrease of the stellar or gaseous surface brightness. This contradicts the well known keplerian-like expectations of the standard Newtonian Gravity if light traces the mass. It is well known that, in order to explain this anomaly, a dark matter (hereafter DM) halo is routinely added in the computation of the gravitational potential \citep{Bosma:1978phd, Bosma:1979aa, Rubin:1980zd}. The luminous and dark matter contributions to the rotation curve (hereafter RC) can explain the observed velocities as traced by star light or gas emission, even at large galactocentric distances where rotational velocities are flat or keep increasing with radius. A recent support for the dark matter particle scenario can be found in \citep{Salucci:2016vxb}.

On the other hand, it is well known that collisionless and stable Weakly Interacting Massive Particles (WIMPs) in a $\Lambda-$Cold Dark Matter (hereafter $\Lambda-$CDM) cosmology, provide a major scenario to frame the DM in galaxies and in the Universe. Cosmological simulations show that the DM density distribution $\rho_{\small{DM}}$ is nearly universal, i.e. dependent only on the mass of the halo. This density profile has a characteristic steep slope (cusp) in the inner regions which can be approximated by a power-law $\rho(r)/\rho_{c}\simeq (r/r_{c})^{-\gamma}$, with $\rho_{c}$ a characteristic constant density, $\gamma$ of order unity, and $r_{c}$ a constant scaling radius.

This cuspy density profile has not been inferred for several dwarfs, spirals and LSB galaxies \citep{Moore:1994yx, Salucci:2000ps, deBlok:2002vgq, deBlok:2008wp, Gentile:2004tb, Gentile:2005de, Oh:2015xoa} whose RCs favor a much shallower cored profile. In addition, independent evidence for a cored distribution comes from stacked RCs which lead to almost solid body profile $V\propto r$ at $r\to0$ \citep{Salucci:2007tm, Donato:2009ab, Karukes:2016eiz}. All this evidence, though remarkable, still needs some investigations because the inner regions of galaxies are dominated by the luminous matter. The cusp-core issue in galaxy DM density profiles is also related to the investigation of the nature of the DM itself and to the galaxy evolution. Numerical simulations have in fact shown that baryonic feedback occurring at later times than the galaxy formation epoch can transform the original NFW profiles into cored ones \citep{Chan:2015tna, DiCintio:2013qxa}. Hence, the existence of cores for DM halo density distributions in dwarf galaxies may not be sufficient to disprove the $\Lambda-$CDM scenario. Specifically, we need for the same galaxy very accurate mass models and a high quality RC which traces, with high resolution, the inner disk kinematics (using mm or optical line tracers) and extends radially for several disk scale lengths (thanks to the outer disk 21-cm line emission of neutral hydrogen).

In this paper we study in detail the DM distribution in M33, a Local Group low-luminosity spiral galaxy, rich in gas and very much dark matter dominated. Its rotation curve is very extended due to the presence of a large gaseous disc and has an excellent spatial resolution due to M33 proximity and to the presence of CO-lines that trace the inner kinematics \citep{Corbelli:2003sn,Corbelli:2014lga}. The high resolution RC of this galaxy gives us the possibility to derive the DM halo distribution from the inner regions out to large galactocentric distances and therefore to uniquely test the $\Lambda-$CDM cosmological scenario. M33 hosts no bulge nor a prominent bar \citep{Corbelli:2007ip}: the absence of these concentrated stellar distributions much alleviates the usual uncertainty in deconvolving the disk contribution from that of a bulge, and in modeling correctly non circular motion. It is important to stress the crucial result obtained for this object by \citet{Corbelli:2014lga}. They have inferred the stellar mass surface density locally by measuring the pixel-SED (Spectral Energy Distribution) and fitting population synthesis models pixel-by-pixel. This method provides a map of the stellar mass density and a good estimate of its radial profile, which previously has been assumed to be proportional to the luminosity one. This reduces considerably the uncertainties on the contribution of luminous matter to the galaxy circular velocity \citep{Corbelli:2014lga}. The gaseous disk mass contribution can be obtained directly from the HI surface photometry given the well known distance of the galaxy.

In \citet{Corbelli:2014lga} the DM density has been derived using the detailed maps of the stellar surface-density in connection with a standard fit (via $\chi^{2}$ minimization) of the observed RC. The mass model considered by the authors, includes a stellar plus a gaseous disk, and a DM halo. The latter is assumed to have a NFW, or alternatively, a cored BRK radial density profile, both with two free parameters to be determined by the RC fit. Their results indicate that both models give a satisfactory RC fit, even though the NFW profile provides a better fit. The DM halo density profile cannot be uniquely determined because the luminous matter significantly contributes to the total gravitational potential in the inner parts of the galaxy. Only for spirals or dwarfs with maximum velocities $<70$ km\,s$^{-1}$ the RC directly measures the DM profile in the inner regions. In objects like M33, instead, even the small uncertainties left in modelling the stellar disk mass, propagate into the inferred DM halo density leaving some uncertainties on its mass and radial density profile.

To better constrain the DM density distribution one can use a second scheme in which the RC and dynamical mass modeling are analyzed only in regions which are not much affected by the luminous matter distribution \citep{Salucci:2010qr}. This method, discussed in \citet{Salucci:2010qr}, has been applied first to the Milky Way and then to the spiral galaxy NGC 3198 \citep{Karukes:2015fma}. It allows to derive very precisely the DM density in the outskirt of a galaxy provided that the HI surface density is well known and the circular velocity and its first derivative are accurately determined, so that $\delta V /V<0.05,$ $\delta d\log V/d\log r <0.1$. Using this method, in this paper we investigate if the M33 RC is compatible with the $\Lambda-$CDM halo density profile or with a cored halo density profile. We have adopted the BRK halo profile to represent the latter, following \citet{Salucci:2007tm, Karukes:2016eiz} who have analyzed the DM halo contribution to the Universal Rotation Curve of spirals.

\bigskip
The paper is organized in four sections. In Sec.\eqref{Sec.2} we summarize the M33 observed rotation curve, the three main baryonic components of the M33 disk, their corresponding surface mass densities and their contributions to the RC. We also outline the results of the previous dynamical analysis \citep{Corbelli:2014lga}. In Sec.\eqref{Sec.3} we develop a local modelling technique that does not rely on any of the global mass modeling of the galaxy to obtain a very robust and careful determination of the DM halo density of M33, and finally, Sec.\eqref{Sec.4} summarizes the main conclusions of this work.

\section{Luminous and dark matter contributions to the rotation curve}\label{Sec.2}

In this section we introduce the stellar and gaseous components of the M33 disk and discuss the radial-dependence of the baryonic surface mass density as derived by \citet{Corbelli:2014lga}. We then briefly summarize how these baryonic mass distributions are used for the standard RC fitting method in M33 and the implications for the DM distribution.

\subsection{Stellar and gaseous disks}

\begin{figure}
\centering
\includegraphics[width=0.6\textwidth]{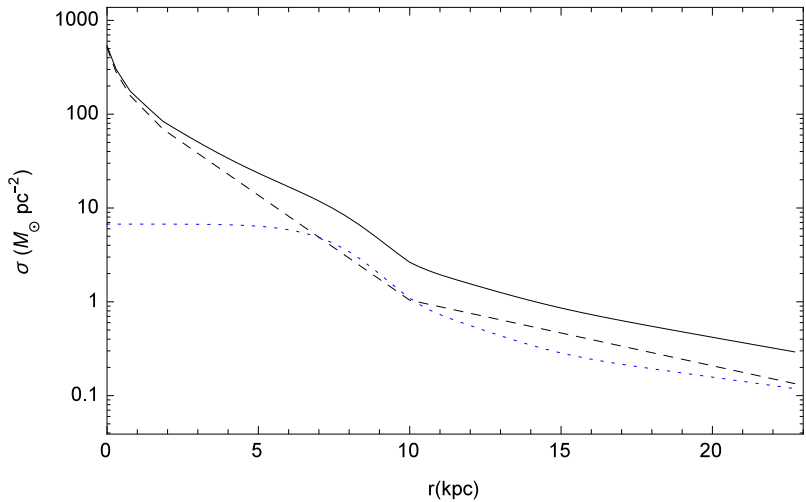}
\caption{Radial dependence of the baryonic mass surface densities of the M33 disk: stellar surface density inferred from BVIgi maps (dashed line), neutral hydrogen surface density (dotted line) and total baryonic mass surface density (continuous line) namely stars, atomic and molecular gas.}
\label{Fig01}
\end{figure}

Thanks to the tight correlations between the colour and the apparent stellar mass-to-light ratio $M/L$ \citep{Bell:2000jt}, the stellar mass, to a first approximation, can be determined using multi-band optical imaging measurements of the whole galaxy luminosity in several bands. However, due to likely radial variations of the stellar mass-to-light ratio, as galaxy disks grow with time. \citet{Portinari:2009ap, Zibetti:2009dp, Delgado:2013ymt} have been using chemo-photometric models for a large samples of spatially resolved, disk-dominated galaxies to determine the radial dependence of the stellar mass surface density. A radially decreasing mass-to-light ratio is found based on galaxy colour gradients and spectral synthesis techniques. Being M33 the second closest spiral, \citet{Corbelli:2014lga} implemented an extension of the ZCR09 method to build up a detailed map of the stellar surface mass density using mosaic maps in the $B$, $V$, $I$, $g$ and $i$ bands from the Local Group Survey \citep{Massey:2006mk} and from the Sloan Digital Sky Survey \citep{York:2000gk}. After processed the images, a pixel-by-pixel synthesis model of the stellar population allowed to obtain a stellar mass surface density map of the M33 disk out to 5~kpc which showed a clear radial gradient of the mass-to-light ratio in the inner regions. This finding is consistent with the negative radial metallicity gradient which supports an inside-out formation scenario and underlines the importance of carrying out a careful analysis of the stellar mass distribution in disks before computing their contribution to rotation curves \citep{Portinari:2009ap}. By fitting the resulting radial averages of the stellar surface mass density of M33, with analytic functions, namely exponential functions with with different scale-lengths $r_{s}$, \citet{Corbelli:2014lga} obtain the following approximation for the radial distribution of the stellar mass surface density in units of $M_{\sun}\,$pc$^{-2}$:
\begin{align}\label{Eq.1}
\sigma_{BVIgi}(r)=
 \begin{cases}
 \exp(-2.010\,r + 6.24), & 0.0\,\text{kpc}\,< r \leq 0.3\,\text{kpc};\\
 \exp(-1.239\,r + 6.01), & 0.3\,\text{kpc}\,< r \leq 0.77\,\text{kpc};\\
 \exp(-0.758\,r + 5.64), & 0.77\,\text{kpc}\,< r \leq 1.85\,\text{kpc};\\
 \exp(-0.515\,r + 5.19), & 1.85\,\text{kpc}\,< r \leq 10 \,\text{kpc};\\
 \exp(-0.161\,r + 1.65), & 10 \,\text{kpc}\,< r \leq 23\,\text{kpc},
 \end{cases}
\end{align}
\noindent where $r$ is the galactocentric radius in kpc units. The fits to the radial distribution of the stellar mass density are shown in Fig.\eqref{Fig01} (dashed line), where we can notice the drop of the stellar mass density by more than 3 orders of magnitudes from the centre to the outskirts of the M33 disk. The amplitude of $\sigma_{BVIgi}(r)$ is uncertain by a $30\%,$ while its radial trend has negligible uncertainties. As in \citet{Corbelli:2014lga}, in order to compute the dynamical contribution of the stellar mass density to the rotation curve, we consider the stellar disk perpendicular to the galactic plane as a flaring disk with a radially varying half thickness: this is only 100~pc at the centre and it reaches 1~kpc at the outer disk edge.

The gaseous disk is made of atomic and molecular gas (mostly hydrogen and helium) and it warps beyond 8~kpc. The high resolution 21-cm data for the atomic hydrogen gas in M33, the best fitting tilted ring model, and the radial averages of the HI surface density distribution, $\sigma_{HI}(r),$ have been presented by \citet{Corbelli:2014lga}. In the same paper, the authors summarize the results of the CO surveys for the determination of the H$_2$ surface mass density. The molecular gas surface density is well represented by the following relation:
$\Sigma_{H2}= 10 \exp{(-r/2.2)}$~M$_\odot$/pc$ ^{-2}$ where $r$ is in kpc \citep{Corbelli:2003sn, Gratier:2010sp, Druard:2014aap}. For the dynamical contribution of the gaseous disk to the rotation curve, the gaseous disk has been considered vertically thick with half thickness of 0.5~kpc.

Using the accurate determination of the stellar mass surface density via chemo-photometric methods, one can determine the total baryonic matter surface density in M33 shown in Fig.\eqref{Fig01}. In the inner regions of the galaxy ($r\lesssim7$ kpc, see Fig.\eqref{Fig01}), the stellar mass surface density dominates over the gaseous mass surface density. For 7 kpc $\lesssim r\lesssim$ 10 kpc, stars and the atomic gas have a similar mass surface density and beyond 10 kpc, they decrease with radius with a similar slope.

In the standard RC fitting method used by \citet{Corbelli:2014lga} the stellar and gaseous disk velocity contributions add in quadrature as follows:
\begin{align}\label{Eq.2}
V_{d}^{2}(r)=\Upsilon V^{2}_{s}(r)+V^{2}_{g}(r),
\end{align}
\noindent where $\Upsilon=\Upsilon_{*}/\Upsilon_{*}^{\text{BVIgi}}=\text{M}_{*}/\text{M}^{\text{BVIgi}}_{*},$ is the ratio between the stellar disk mass and that given by Eq.\eqref{Eq.1}, which is 4.9$\times 10^9$~M$_\odot$ out to $r=23$~kpc. The parameter $\Upsilon$ takes into account the uncertainties in the stellar disk mass and hence, the total stellar disk mass is allowed to vary in the interval $3.4-6.4\times 10^9$~M$_\odot$. The terms $V_{s}(r)$ and $V_{g}(r)$ indicate the contributions to the rotation curve of the stellar disk, as given by of Eq.\eqref{Eq.1}, and of the Helium corrected HI and H$_{2}$ gas mass.

\subsection{Dark matter halo models and dynamical analysis of the rotation curve}

The standard dynamical analysis of the RC is carried out by modeling the galaxy with a DM halo and a stellar and gaseous disk mass surface densities in the whole sampled radial range ($0.24\leq r\leq 23$ kpc for M33). Two different DM halo models have been considered by \citet{Corbelli:2014lga} in fitting the M33 RC. The first one is a halo with a NFW density profile for structures growing in a $\Lambda-$CDM hierarchical universe \citep{Navarro:1995iw, Navarro:1996gj, Klypin:2000hk, Hayashi:2004vm, Diemand:2005wv,Caimmi:2004qj} given by:
\begin{align}
\rho_{\small{NFW}}(r)=\dfrac{\rho_{c}}{\frac{r}{r_{c}}\left(1+\frac{r}{r_{c}}\right)^{2}},\label{Eq.3}
\end{align}
\noindent where $\rho_{c}$ and $r_{c}$ are the two usual halo parameters. The contribution of the DM density profile adds in quadrature to the disk contribution to give the total rotational velocity as
\begin{align}
&V^{2}(r)=V_{\small{NFW}}^{2}(r)+V_{d}^{2}(r),\label{Eq.4}\\
&V_{\small{NFW}}^{2}(r)=\dfrac{4\pi G\rho_{c}r_{c}^{2}}{r/r_{c}}\left(\ln(1+r/r_{c})-\dfrac{r/r_{c}}{1+r/r_{c}}\right).\label{Eq.5}
\end{align}

The virial mass $\text{M}_{\text{vir}}$ and the concentration parameter $c=r_{vir}/r_{c},$ are related to $\rho_{c}$ and $r_{c}$ by:
\begin{align}\label{Eq.6}
&\rho_{c}=\dfrac{97.2}{3}\dfrac{c^{3}}{\ln(c+1)-\frac{c}{c+1}}\rho_{\text{crit}}\;\;\text{g cm}^{-3},\\
&r_{c}=\dfrac{1}{c}\left(\dfrac{3}{97.2}\dfrac{\text{M}_{\text{vir}}}{4\pi\rho_{\text{crit}}}\right)^{1/3}\;\;\text{kpc}.
\end{align}
\noindent where $r_{\text{vir}}$ is the virial radius and $\rho_{\text{crit}}=9.3\times10^{-30}$ g\,cm$^{-3}$ is the critical density of the Universe. The free parameters, $c$ and $\text{M}_{\text{vir}}$ have been determined as independent parameters, but after the RC fit, the authors checked the consistency of their values with the relations obtained by numerical simulations of structure formation in a $\Lambda-$CDM universe.
The free parameters of this halo model, $\text{M}_{\text{vir}}$ and $c,$ are in fact not independents from each other \citep{Bullock:1999he}. Recent numerical simulations \citep{Dutton:2014xda} suggest a correlation which can be expressed using the dimensionless value of the Hubble parameter $h=0.678$ \citep{Ade:2015xua} as:
\begin{align}\label{Eq.7}
c=10.6\left(\dfrac{1}{10^{12}h^{-1}}\dfrac{\text{M}_{\text{vir}}}{\text{M}_{\odot}}\right)^{-0.097}.
\end{align}

The following values have been obtained by \citet{Corbelli:2014lga}: $c=(9.5\pm1.5),$ $\text{M}_{\text{vir}}=(4.3\pm 1.0)\times10^{11}\;\text{M}_{\sun},$ and $\text{M}_{*}=(4.9\pm0.6)\times10^{9}$ M$_{\sun}$ by considering a composite probability, which takes into account the fit to the RC, the synthesis models of the stellar population and the $c-$M$_\text{vir}$ relation found by numerical simulations.

The second halo profile considered by \citet{Corbelli:2014lga} is the phenomenological cored BRK density distribution, that successfully fits individual RCs (e.g. \citet{Gentile:2005de} and references therein), as well as the Universal Rotation Curve of spirals. This cored halo profile, also known as the Burkert halo profile \citep{Burkert:1995yz, Salucci:2000ps} has a radial density distribution given by
\begin{align}\label{Eq.8}
\rho_{\small{BRK}}(r)=\dfrac{\rho_{c}}{\left(1+\frac{r}{r_{c}}\right)\left(1+\frac{r^{2}}{r_{c}^{2}}\right)}
\end{align}
\noindent where $\rho_{c}$ and $r_{c}$ are the central density and the core radius respectively. This density profile induces a gravitational potential such that every particles in the galactic disk rotates with a velocity $V_{\small{BRK}}$ which adds in quadrature to the velocity induced by the baryonic disk component as:
\begin{align}
&V^{2}(r)=V_{\small{BRK}}^{2}(r)+V_{d}^{2}(r),\label{Eq.9}\\
&V_{\small{BRK}}^{2}(r)=\dfrac{2\pi G\rho_{c}r_{c}^{2}}{r/r_{c}}\left(\ln(1+r/r_{c})+\ln\sqrt{1+(r/r_{c})^{2}}- tan^{-1}(r/r_{c})\right).\label{Eq.10}
\end{align}

The central core density $\rho_{c}$ and the scaling radius $r_{c}$ are not independents either, as shown in \citet{Donato:2009ab}, and the following relation holds:
\begin{align}\label{Eq.11}
\log\left(\dfrac{\rho_{c}r_{c}}{\text{M}_{\sun}\text{pc}^{-2}}\right)=2.15\pm0.2.
\end{align}

Through the dynamical analysis of the M33 RC, \citet{Corbelli:2014lga} found no compatible values of the stellar mass surface densities using Eq.\eqref{Eq.1}. However, considering only the $B,V,I$ color maps of M33, the synthesis models predict somewhat higher stellar masses than Eq.\eqref{Eq.1}: 5.5$\times 10^9$~M$_\odot$. Given the uncertainties in the models, the likely M33 stellar mass is in this case within the interval 3.9-7.2$\times 10^{9}$~M$_\odot$. Using the corresponding stellar mass surface density the following RC best fitting values for the BRK halo and the stellar mass have been found: $r_{c}=7.5$ kpc, $\rho_{c}= 18.0\times10^{6}$ M$_{\sun}$ kpc$^{-3}$ and $\text{M}_{*}=7.2\times10^{9}$ M$_{\sun}$. The global $\chi^{2}$ value is larger than that found when using the NFW DM profile but still acceptable. Therefore, even for this galaxy, despite we have accurate and extended RC data tracing the gravitational potential, and a good determination of the baryonic mass surface density, both DM mass models are compatible with the data. This can be easily attributed to the fact that in M33, even beyond $10$ kpc, the influence of the luminous matter is not negligible so that any derivation of the DM density profile would be fraught with the uncertainties inherent to such a component. More precisely, since in the inner regions $r\lesssim 2R_{D}$ (being $R_{D}$ the disk length scale) of most spirals, the stellar disk dominates over the dark component, even small uncertainties in the mass determination of the former induces large uncertainties in the values of the structural parameters of the dark components. By using a second method for fitting the RC data in the next section we hope to alleviate this degeneracy. We will see that the fit related to the second method is less dependent on the luminous matter distribution and thus hopefully it will help us to clarify better the properties of the DM halo that hosts M33.

\section{Model-independent method for local density estimation.}\label{Sec.3}


A new method to determine the DM density distribution in spiral galaxies has been introduced by \citet{Salucci:2010qr}. This method was applied first to estimate the value of the DM density at the Sun's location \citep{Salucci:2010qr}, and extended in \citet{Karukes:2015fma} to the study of the DM distribution in the galaxy NGC 3198. The goal of this section is to derive, for the spiral galaxy M33, a model independent DM density using this method which deals with the RC at large radii, where the influences of the stellar and gaseous disks are weak.

\subsection{The local density estimation method}

The idea, brought by \cite{Salucci:2010qr} is to resort the equation of centrifugal equilibrium, which holds the spiral galaxies:
\begin{align}\label{Eq.12}
\dfrac{V^{2}}{r}=a_{h}+a_{s}+a_{g},
\end{align}
\noindent where $a_{h}, a_{s},$ and $a_{g}$ are the radial accelerations, generated by the DM halo, stellar and gaseous disks respectively.

Under the approximation of a spherical DM halo, we have
\begin{align}
\rho_{h}(r)&=\rho(r)-\Upsilon\rho_{s}(r)-\rho_{g}(r)=\dfrac{X_{q}}{4\pi G r^{2}}\dfrac{d}{dr}\left(rV^{2}-r\Upsilon V^{2}_{s}-rV^{2}_{g}\right),\label{Eq.13}\\
\rho(r) &=\dfrac{1}{4\pi G r^{2}}\dfrac{d}{dr}\left(rV^{2}\right),\label{Eq.14}\\
\rho_{s}(r)&=\dfrac{1}{4\pi G r^{2}}\dfrac{d}{dr}\left(rV^{2}_{s}\right),\label{Eq.15}\\
\rho_{g}(r)&=\dfrac{1}{4\pi G r^{2}}\dfrac{d}{dr}\left(rV^{2}_{g}\right),\label{Eq.16}
\end{align}
\noindent where $X_{q}$ is a factor correcting the spherical Gauss law used above in case of an oblate DM halo and it takes values between 1.05 and 1.00 (see details in \citet{Salucci:2010qr}), $V(r)$ is the velocity given by the RC, $V_{s}$ and $V_{g}$ are the stellar and gas velocities and $\Upsilon$ the usual stellar mass-to-light ratio. The strength of this method lies in the fact that we have transformed the surface mass density of the stellar and gaseous disks in equivalent bulk densities with the aid of Gauss's law. Since the velocities induced by the stellar and gaseous disks decrease slowly after $9.53$~kpc, we expect a sharp fall for their equivalent three-dimensional densities $\rho_{s}$ and $\rho_{g},$ and we are just left only with the DM contribution to the observed RC. From this fact, we can infer the DM halo properties directly from the experimental data.

\bigskip
An issue left to treat is to derive an analytical expression for the total velocity $V(r)$ from the available discreet set of the RC data. Having an analytic function will avoid artifact due to the numerical computation of the derivative of $V(r)$ needed for the computation of the total density as given by Eq.\eqref{Eq.14}. The next subsection is devoted to find an appropriate empirical smooth curve to solve this problem.

\subsection{Empirical velocity profile}

With the aid of the local density method, we can study, directly from the experimental data, the DM halo properties by computing the densities given by Eq.\eqref{Eq.14}, Eq.\eqref{Eq.15} and Eq.\eqref{Eq.16} respectively. Just with the purpose of deriving a smooth profile of $dV/dr$, we introduce the following an empirical velocity formula to describe the rotation curve of M33 :
\begin{align}\label{Eq.17}
V(r)=V_{0}\dfrac{r/r_{0}+d}{r/r_{0}+1}.
\end{align}
\noindent

This is a simple, rational, 3-parameter-dependent velocity profile that reproduces at large galactocentric distances a flat velocity curve and as we will see immediately, it describes genuinely the RC in the range allowed by the experimental data. The three free parameters, namely the terminal velocity $V_{0}$, the scaling radius $r_{0}$ and $d<1$, an additional parameter to prevent $V(r)$ to be constant, make the degree of freedom of the fit to be similar to that of the standard RC fitting method (see previous Section) when considering the uncertainties on the stellar surface mass density. The empirical velocity formula can be used when considering both for the BRK halo plus luminous disk and the NFW halo plus luminous disk mass models.

\begin{figure}
\centering
\includegraphics[width=0.45\textwidth]{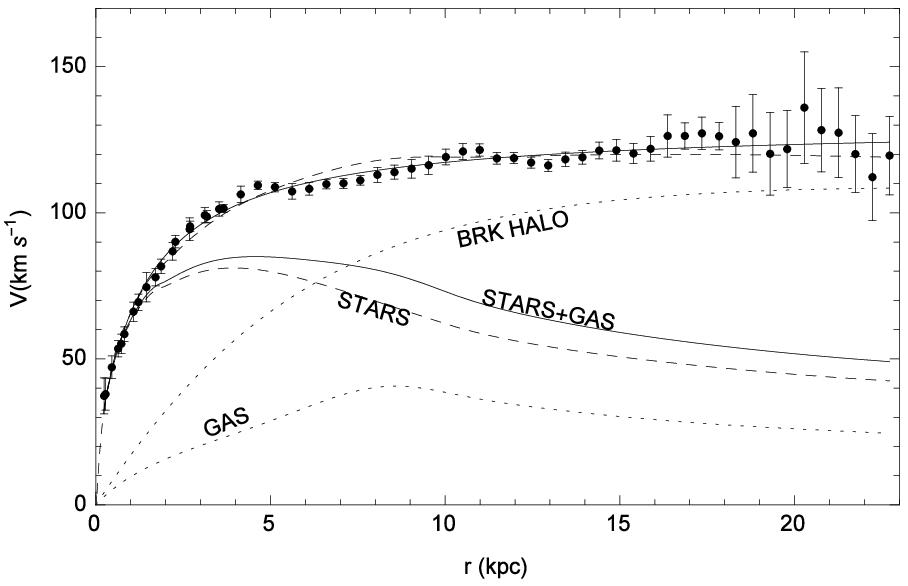}
\includegraphics[width=0.45\textwidth]{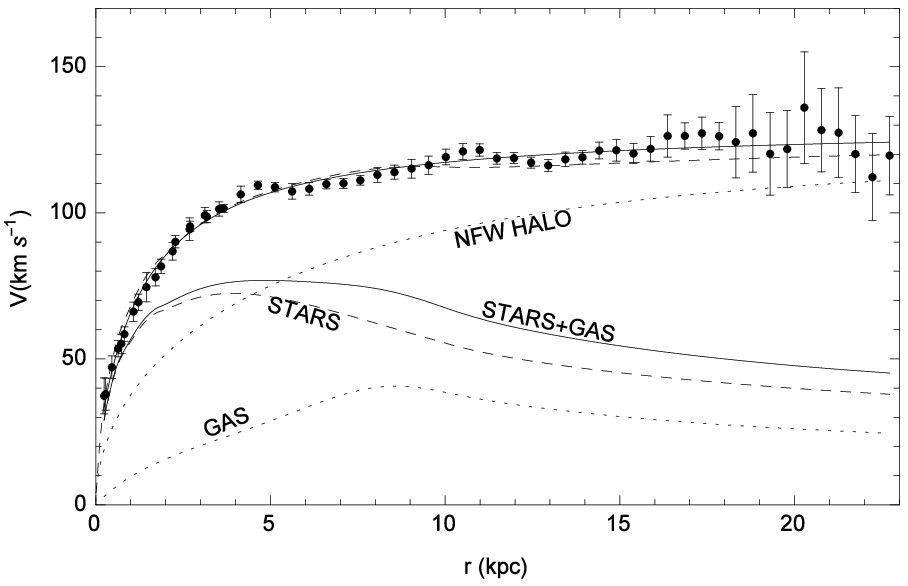}
\caption{RC of M33 (black dots with error bars) with the analytical best fit to it as given by Eq.\eqref{Eq.17} (continuous line) with $V_{0}$ = 130.2 ~km~s$^{-1}$, $r_0$ = 1.3~kpc, and d = 0.12. In both panels are also shown (in dashed lines with no labels), for comparison, the solutions to the RC fit obtained by \citet{Corbelli:2014lga} using Eq.\eqref{Eq.9} (left panel) and Eq.\eqref{Eq.4} (right panel) respectively, as discussed in the previous section. The halo, stellar, gas and total disk contributions are shown as well with their corresponding labels.}
\label{Fig02}
\end{figure}

\begin{figure}
\centering
\includegraphics[width=0.315\textwidth]{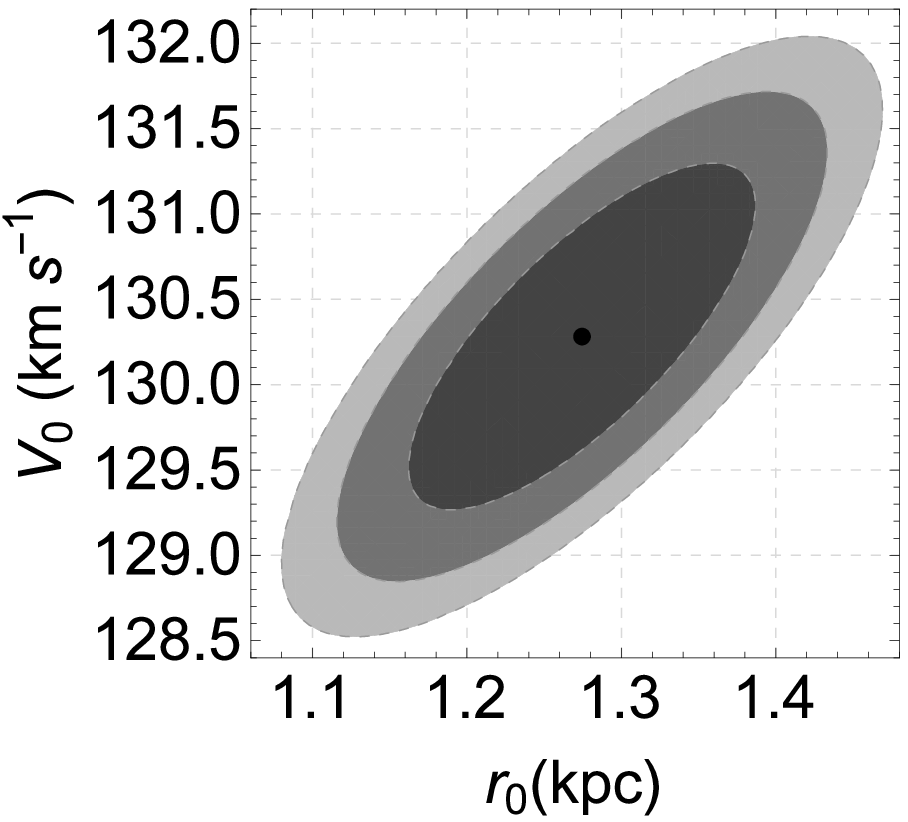}
\includegraphics[width=0.3\textwidth]{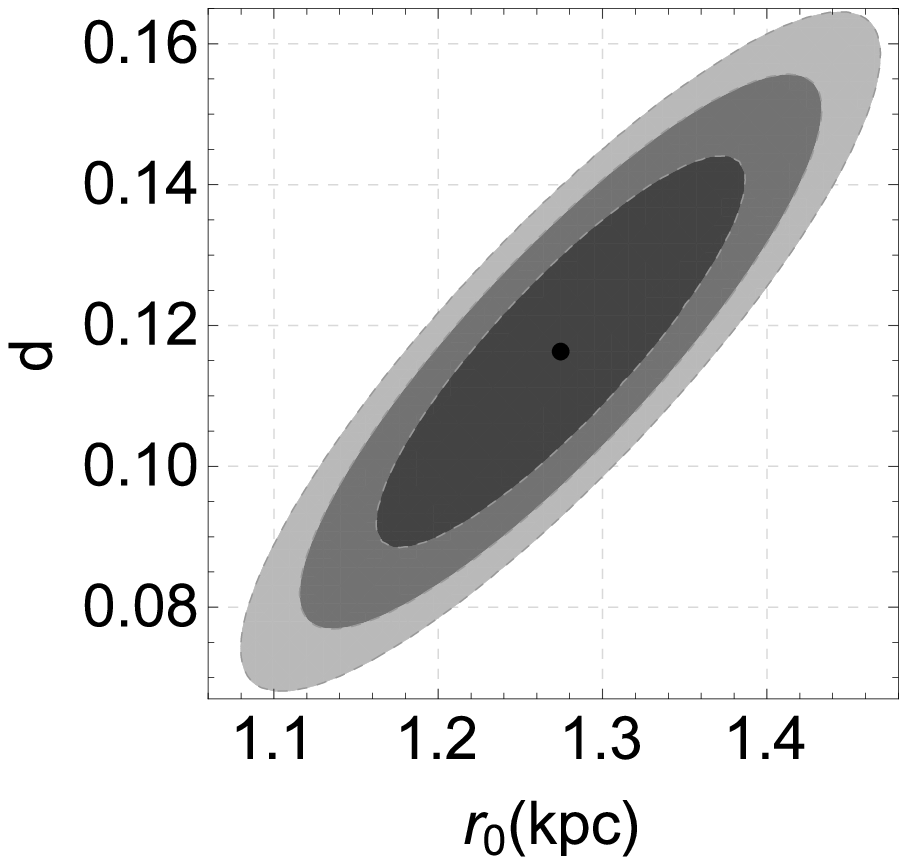}
\includegraphics[width=0.33\textwidth]{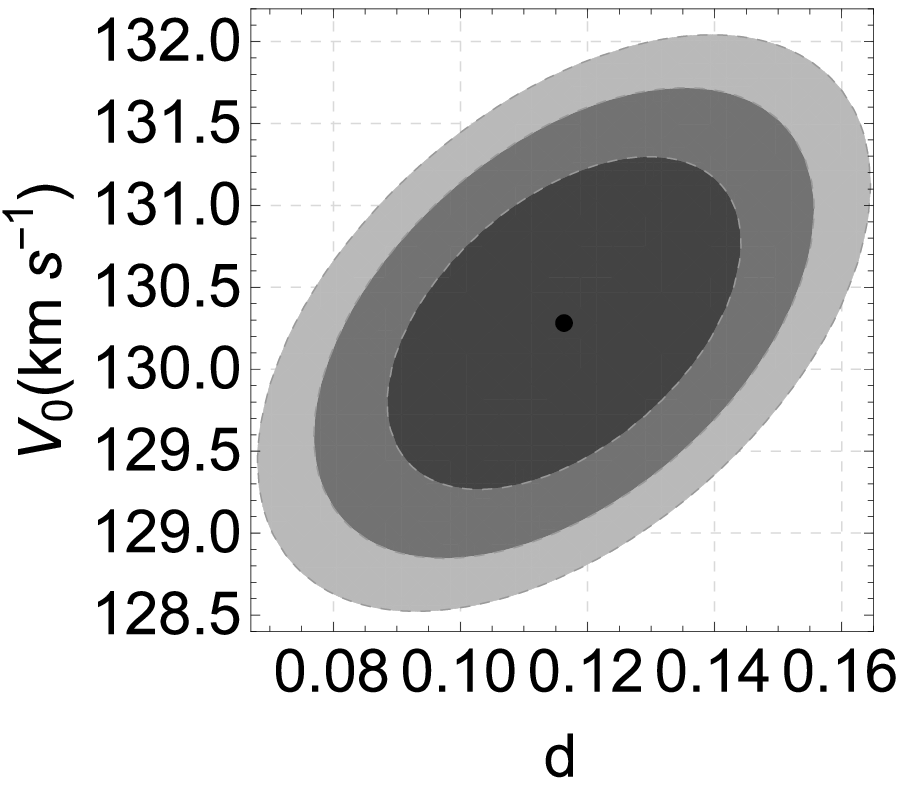}\\
\caption{1, 2 and 3$\sigma$ confidence ellipses (from dark to light) for the best fitting parameters $V_{0},\;r_{0},\;d$ in fitting the analytic function in Eq.\eqref{Eq.17} to RC data.
The black central dots indicate the best fitting values.}
\label{Fig03}
\end{figure}

By minimizing the $\chi^{2}$ distribution for fitting the RC data with its measurement errors, with the above analytic function, we obtain the best fit parameters: $V_{0}=(130.2\pm1.0)$ km s$^{-1},$ $r_{0}=(1.3\pm0.1)$ kpc, and $d=(0.12\pm0.03),$ giving a $\chi_{\text{red}}^{2}=0.75$. In Fig.\eqref{Fig02} we show the fit to the RC data of the derived analytic function. For comparison we show also the fits obtained by \cite{Corbelli:2014lga} using the BRK DM profile (left panel) and the NFW DM profile (right panel) discussed in the previous section. Moreover, in Figs.\eqref{Fig03} we show the $1,2,3\sigma$ confidence levels for such three free parameters.

\subsection{Dark matter profiles and effective densities}

Given the goodness of the fit of the RC data with the analytic function given by Eq.\eqref{Eq.17}, the goal of this subsection is to use this continuous smooth curve to compute $\rho(r)$ as given by Eq.\eqref{Eq.14} and the effective baryonic disk densities. As mentioned before, for $r\gtrsim 9$ kpc the contributions of the star and gas densities are negligible in Eq.\eqref{Eq.13} and we are left only with the DM halo contribution. 

\begin{figure}
\centering
\includegraphics[width=0.576\textwidth]{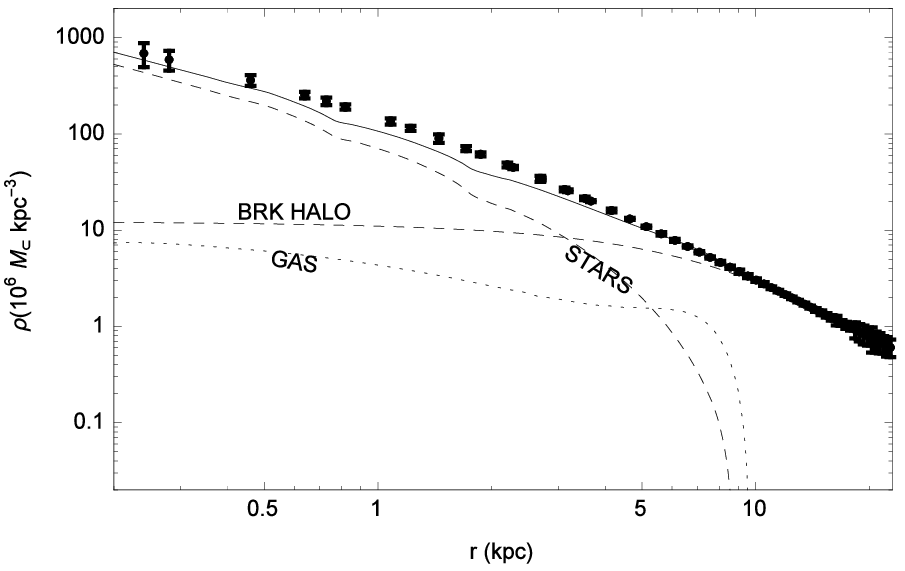}
\includegraphics[width=0.364\textwidth]{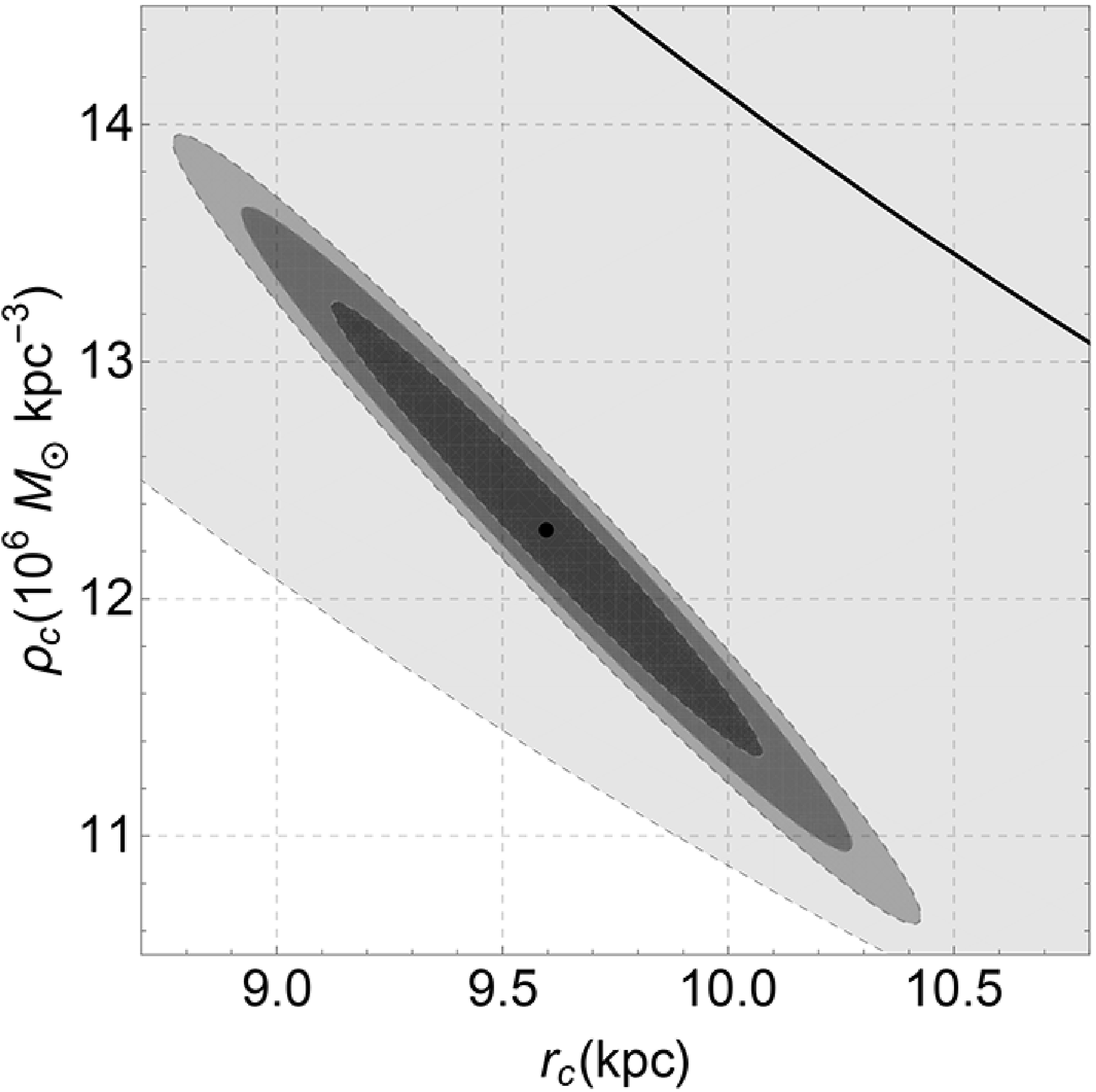}
\caption{The left panel shows the dependence of various effective densities on galactocentric radius: the black points with error bars correspond to $\rho(r)$ as derived from Eq.\eqref{Eq.14}. The continuous line marks the total density given by the BRK halo plus stars and gas. We also plot the effective densities of stars and gas given by Eq.\eqref{Eq.15} and Eq.\eqref{Eq.16} respectively and the best fitting BRK DM halo density. The right panel shows the 1, 2 and 3$\sigma$ confidence ellipses (from dark to light) for the best fitting parameters $r_{c}$ and $\rho_{c}$, as well as the correlation relation between these two (continuous line) given by Eq.\eqref{Eq.11} with a $20\%$ of uncertainty (shaded area).}
\label{Fig04}
\end{figure}

In the range of galactocentric distances $9.5\,\text{kpc}\leq r\leq22.72\,\text{kpc},$ we fit the derived density profile using the BRK and NFW DM radial density distribution. In the case of the BRK profile, we obtain the best fitting values: $r_{c}=(9.6\pm0.5)$ kpc and $\rho_{c}=(12.3\pm1.0)\times10^{6}$ M$_{\sun}$ kpc$^{-3}$ respectively, giving a $\chi_{\text{red}}^{2}=0.8$ and the halo virial mass is $\text{M}_{\text{vir}}(3.0\pm0.8)\times10^{11}\,\text{M}_{\sun}$. In Fig.\eqref{Fig04} (left panel) we show the corresponding fit in log-log scale. Framed by this solution, let us stress again that close to $r \sim 9-9.5$~kpc the stars and gas contributions drop sharply more than two orders of magnitude, therefore, in the radial range $9.53\,\text{kpc}\leq r\leq22.72\,\text{kpc}$ the stellar mass-to-light ratio $\Upsilon$ plays no role in this analysis and we can obtain the two halo parameters also if the baryonic disk mass has some uncertainties. In the right panel of Fig.\eqref{Fig04} we show the corresponding $1,2,3\sigma$ confidence levels for the two BRK halo parameters. In the same panel, the continuous line shows the correlation found (Eq.\eqref{Eq.11}) and the shaded area its $1\sigma$ region. Clearly the solution obtained for the parameters $\rho_{c}$ and $r_{c}$ lies inside the uncertainties associated with the correlation, as reported in \citet{Donato:2009ab}.

We performed the same analysis using the NFW profile. In this case we obtained the best fitting values $c=(9.5\pm0.7)$ and $\text{M}_{\text{vir}}=(5.4\pm0.6)\times10^{11}$ M$_{\sun},$ giving a $\chi_{\text{red}}^{2}=1.0.$ In the left panel of Fig.\eqref{Fig05} we show the corresponding fit in log-log scale. In the right panel we display the corresponding $1,2,3\sigma$ confidence levels for the two free parameters of the NFW profile, as well as their correlation relation found by numerical simulations and its uncertainties as given by Eq.\eqref{Eq.7} (shaded area).

\begin{figure}
\centering
\includegraphics[width=0.576\textwidth]{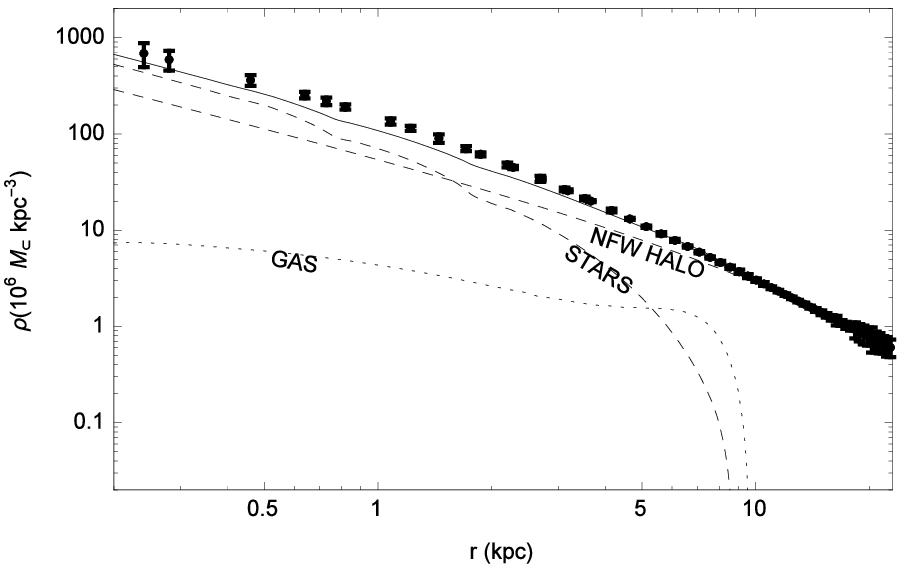}
\includegraphics[width=0.38\textwidth]{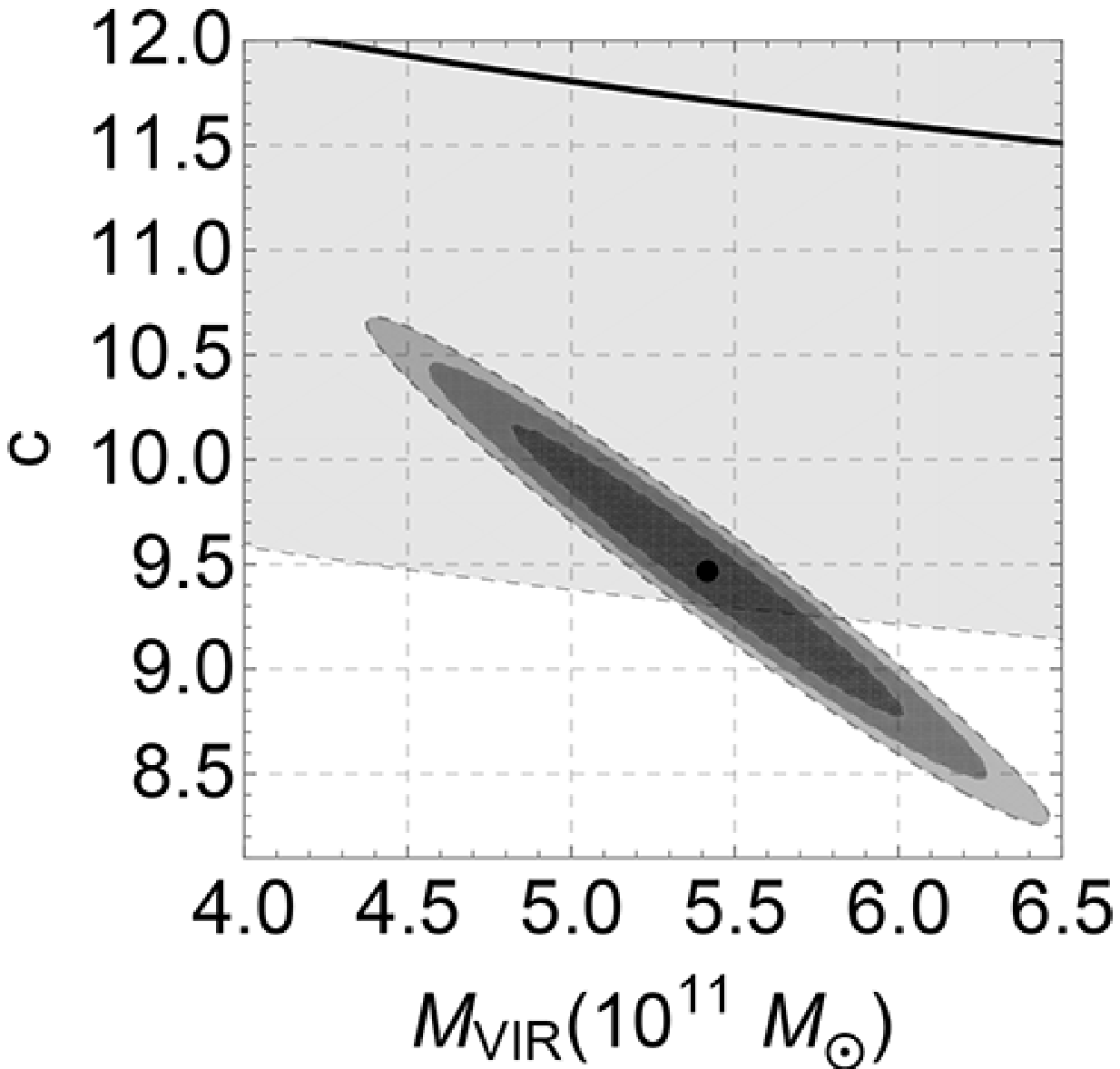}
\caption{The left panel shows the dependence of various effective densities on galactocentric radius: the black points with error bars correspond to $\rho(r)$ as derived from Eq.\eqref{Eq.14}. The continuous line marks the total density given by the NFW halo plus stars and gas. We also plot the effective densities of stars and gas given by Eq.\eqref{Eq.15} and Eq.\eqref{Eq.16} respectively and the best fitting NFW DM halo density. The right panel shows the 1, 2 and 3$\sigma$ confidence ellipses for the best fitting halo parameters $c$ and $\text{M}_{\text{vir}}$, where the black central point corresponds to the best fitting values of $c$ and $\text{M}_{\text{vir}}$. The continuous line and shaded area in the right panel correspond to the $c-\text{M}_{\text{vir}}$ correlation given by Eq.\eqref{Eq.7}.}
\label{Fig05}
\end{figure}

It is worth to discuss the present result. We can trace the distribution of matter in M33 using the RC observational data starting from the innermost radius $r=0.24$ kpc. Inside this radius, no information can be inferred except by extrapolating the best fitting models and testing the predicted RC velocities with follow up higher resolution observations. In the radial range $0.24$ kpc $\leq r\lesssim 9$ kpc, the stellar component is dynamically very relevant with respect to the DM halo component. For both DM halo models tested, the structural parameters of the DM distribution are mingled with the unknown value of the disk mass. In both cases, no information can be extracted either since any measurements of DM distribution would be fraught with the uncertainties due to the stellar contribution (i.e. the stellar disk accounts for most of the total gravitational potential of the galaxy). Around $9$ kpc, the luminous matter still competes against DM to dominate over the total mass, given the RC uncertainties. But beyond 9.5~kpc, at $\simeq3.5 R_{D},$ the exponential decrease of the stellar matter density and the negligible contribution of the gaseous component, which never plays a role even if its total mass is about half of that of the stars, assure that only the DM component balances the radial accelerations, which the gaseous disk is engaged with.

In Fig.\eqref{Fig05} we show the best fits of the BRK and NFW profiles to the outer disk effective matter densities. As we can notice, the values and radial gradients of the matter density of the two DM halo models are similar, in agreement with our finding that both halos are compatible with the data. This is because the core of the BRK best fitted halo does not extend much beyond the
optical disk, where DM is the dominant dynamical component.

\begin{figure}
\centering
\includegraphics[width=0.6\textwidth]{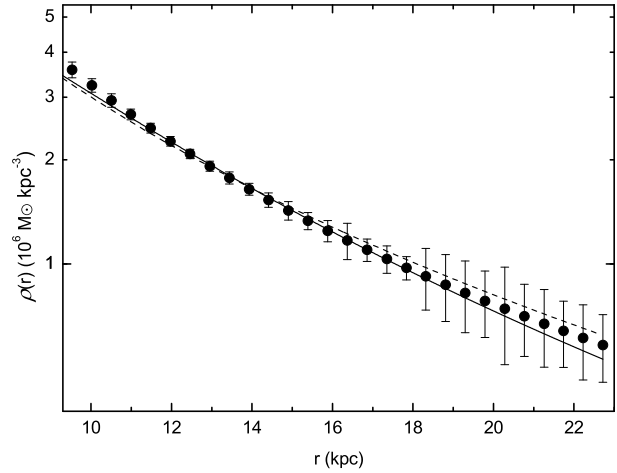}\\
\caption{The observed DM halo density profile in the outer disk is shown using black dots with error bars. For comparison the BRK (continuous line) and NFW (dashed line) best fitting halo models to the effective densities inferred from the dynamical analysis of RC data are also displayed.
}
\label{Fig05}
\end{figure}

\section{Conclusions}\label{Sec.4}

In this paper we deepened the results of \citet{Corbelli:2014lga} which obtained a global fit of the RC of M33, after an accurate mass modelling of its stellar disk, by considering a DM halo with a cuspy NFW or a cored BRK profile. The results of the global fit to the RC favour a cuspy $\Lambda-$CDM halos despite a cored one is still compatible with the data when a heavy stellar disk is considered. Following \citet{Salucci:2010qr} and \citet{Karukes:2015fma} we then use a new approach to investigate the DM content of M33. Relying on the centrifugal equilibrium condition, assuming a spherical DM halo, we trace the DM density by using only the experimental RC. This method relies on fitting the dark matter density distribution of M33 in the radial range $9.53\,\text{kpc}\leq r\leq22.72\,\text{kpc},$ where the stellar and gaseous contributions to the RC are negligible. According to this method, the BRK profile with the following parameters: $r_{c}=(9.6\pm0.5)$ kpc and $\rho_{c}=(12.3\pm1.0)\times10^{6}$ M$_{\sun}$ kpc$^{-3}$ provides an excellent fit to the M33 data. The BRK halo virial mass is $\text{M}_{\text{vir}}=(3.0\pm0.8)\times10^{11}\,\text{M}_{\sun}$ and it is only $6.7\pm1.2\times10^{10} \text{M}_{\sun}$ out to about 23 kpc. When testing the NFW DM profile through the local density estimator method we find instead a higher virial mass for the halo and therefore an extremely low baryonic fraction. With the analysis presented in this paper, we conclude that the possibility that M33 harbors a cored dark matter halo seems still suitable. Even though the determination of the stellar disk mass via synthesis models has eliminated most of the disk-halo degeneracies in this low luminosity spiral, we have shown that the cuspy or cored dark matter density profiles degeneracy still hold because different dynamical methods used in modelling RC data favour different dark matter density distributions and halo masses.



\bsp	
\label{lastpage}
\end{document}